# Cavity-enhanced induced coherence without induced emission

MINHAENG CHO[1,2,*] AND PETER W. MILONNI[3]

[1]Center for Molecular Spectroscopy and Dynamics, Institute for Basic Science (IBS), Seoul 02841, Republic of Korea.

[2]Department of Chemistry, Korea University, Seoul 02841, Republic of Korea.

[3]Department of Physics and Astronomy, University of Rochester, Rochester, New York 14627, USA

* mcho@korea.ac.kr

**Abstract.** This paper presents a theoretical study of the enhancement of Zou-Wang-Mandel (ZWM) interferometry through cavity-enhanced spontaneous parametric down-conversion (SPDC) processes producing frequency-entangled biphotons. The ZWM interferometry shows the capability to generate interference effects between single signal photons via indistinguishability between the entangled idler photons. This paper extends the foundational principles of ZWM interferometry by integrating cavity-enhanced SPDCs, aiming to narrow photon bandwidths for improved coherence and photon pair generation efficiency, which is critical for applications in quantum information technologies, quantum encryption, and quantum imaging. This work explores the theoretical implication of employing singly resonant optical parametric oscillators within the ZWM interferometer to produce narrow-band single photons. By combining cavity-enhanced SPDCs with ZWM interferometry, this study fills a gap in current theoretical proposals, offering significant advancements in quantum cryptography and network applications that require reliable, narrow-band single photons.

# 1. INTRODUCTION

Zou-Wang-Mandel interferometry, often characterized as "induced coherence without induced emission," represents a remarkable advancement in the field of quantum optics and interferometry.[1, 2] This concept was first introduced by Zou, Wang, and Mandel (ZWM) in the early 1990s, marking a significant departure from classical interpretations of light behavior and coherence.

This interferometry involves two spontaneous down-conversion (SPDC) processes occurring in separate second-order nonlinear crystals. Each SPDC is induced by a classical pump beam, which results in the generation of signal and idler photons that satisfy the conservation of energy, i.e., $\omega_P = \omega_S + \omega_I$, where $\omega_P$, $\omega_S$, and $\omega_I$ are, respectively, the angular frequencies of the pump, signal, and idler beams.[3] The two SPDC processes are specifically arranged such that the idler photons from the two crystals are indistinguishable. Interestingly, despite the absence of a direct optical path between the two signal fields generated in each process, the setup can produce interference effects.[2, 4] This phenomenon is counterintuitive because it suggests the establishment of coherence between two light (signal) beams without the necessity of photon emission from one beam to the other, hence the term "induced coherence without induced emission."

ZWM interferometry underscores the subtle and non-intuitive nature of quantum mechanics, particularly in the context of quantum coherence and entanglement.[4, 5] It serves as a foundation for developing new technologies and for a better understanding of the quantum theory of light, bridging the gap between theoretical quantum physics and practical applications in optical sciences.

The implications of this discovery are profound, extending beyond mere academic curiosity to practical applications in quantum imaging with undetected photons, quantum information, quantum spectroscopy, and precision measurements.[6-13] For instance, it provides a method to enhance the resolution of imaging systems beyond classical limits and offers novel approaches to quantum entanglement and information transfer.[14, 15]

SPDC generates pairs of photons exhibiting entanglement across various properties. The pair, composed of a signal photon and an idler photon, is often referred to as a biphoton. By selecting specific idler photons, this method can be utilized to produce single photons on demand, which are crucial for numerous quantum information technologies, including quantum encryption and quantum computing utilizing linear optics.

Interesting proposals for quantum networks envision utilizing stationary quantum bits, like atoms or ions, as nodes for processing information and single photons for information transmission through optical fibers.[16-19] The initial components of such a network have been successfully implemented.[20, 21] To effectively link photons with atomic transitions, the photons' bandwidth must be significantly narrowed to match the atomic transition's linewidth, which is much narrower than the bandwidth of spontaneously emitted photons in SPDC.[22, 23] Furthermore, for quantum networks to function reliably, a consistent photon output is necessary, which requires active stabilization of the OPO.

To narrow the photon bandwidth, one can employ cavity-enhanced PDC.[24] Indeed, this technique was utilized to create a robust source of heralded narrow-band single photons using a singly or doubly-resonant optical parametric oscillator (OPO). The OPO contained a nonlinear crystal within a cavity that resonated with either the signal photon or both signal and idler photons. An external filter then selected narrow-band single signal photons. To specifically produce single photons, the OPO must operate well below its threshold, ensuring that the rate of photon pair production is lower than the cavity's loss rate.

Prior experiments on cavity-enhanced SPDC have achieved biphoton production using an OPO under these conditions.[25-32] The initial theoretical framework, based on the input-output theory developed by Collette and Gardiner [29] to describe the squeezing of intracavity modes in parametric amplification, was already presented by Herzog et al.[33] for a singly-resonant OPO and by Scholz et al.[34] for a doubly-resonant OPO to describe the single photon states produced.

To date, a theoretical proposal or study of cavity-enhanced ZWM interferometry, where the SPDC process inside an optical cavity still operates below the threshold, has been lacking. This study aims to combine the cavity-enhancement effect with the two SPDC processes nested in the ZWM interferometer. This cavity-enhanced induced coherence without induced emission might be used for quantum cryptography and network applications utilizing narrow-band single photons.

## 2. RESULTS AND DISCUSSION

### A. Cavity-enhanced SPDC

The ZWM interferometer requires two SPDC crystals.[2] Here, we consider type-II SPDC with a nonlinear crystal of length $l$. The polarization of the pump is parallel (perpendicular) to that of the signal (idler) field. In this section, we first consider a single SPDC in a cavity resonant with the signal field but not with the idler (Figure 1(a)). In addition to energy conservation, momentum conservation requires $\boldsymbol{k}_P(\omega_P) = \boldsymbol{k}_S(\omega_S) + \boldsymbol{k}_I(\omega_I)$, where $\boldsymbol{k}$ is the corresponding wave vector. The PDC-generated signal field leaves the cavity in a positive $z$ direction, and, due to cavity resonance, there are standing waves composed of two components propagating in the negative and positive $z$ directions.

Treating the pump field classically and the signal and idler fields quantum mechanically, we have the field-matter interaction Hamiltonian for the PDC process, i.e.,

$$\widehat{H}_{int} = \frac{\chi}{2l}\int_{-l}^{0}dx\left(E_{P,c}\widehat{E}_{S,c}^{(-)}\widehat{E}_{I,c}^{(-)} + E_{P,c}^{*}\widehat{E}_{S,c}^{(+)}\widehat{E}_{I,c}^{(+)}\right), \tag{1}$$

where $\chi$ is the second-order susceptibility, $E_{P,c}$ is the monochromatic pump electric field ($=E_P e^{i[k_P(\omega_P)z-\omega_P t]}$), and $\widehat{E}_{m,c}^{(+)}$ ($m = S$ or $I$) represents the positive-frequency electric field operator in the PDC crystal. Here, using the input-output theory for the cavity,[33, 35] we have

$$\widehat{E}_{S,c}^{(+)}(z,t) = \sqrt{\frac{\hbar\omega_S}{\epsilon_0 n_S cA}}\frac{\sqrt{\Delta\omega_c}}{2\pi}\sum_{m=-\infty}^{\infty}\int_{-\infty}^{\infty}d\Omega\frac{\sqrt{\gamma}}{\frac{\gamma}{2}+i\Omega}\hat{a}(\omega_m+\Omega)e^{i[k_{S,m}(\Omega)z-(\omega_m+\Omega)t]}$$

$$\widehat{E}_{I,c}^{(+)}(z,t) = \sqrt{\frac{\hbar\omega_I}{2\epsilon_0 n_I cA}}\int_{-\infty}^{\infty}\frac{d\Omega}{\sqrt{2\pi}}\hat{b}(\omega_I+\Omega)e^{i[k_I(\Omega)z-(\omega_I+\Omega)t]}, \tag{2}$$

where $\omega_m$ are the angular frequencies of the cavity-resonant signal modes (standing waves) that are approximately given by $\omega_m \cong \omega_S + m\Delta\omega_c$ with $\Delta\omega_c$ being the free spectral range inside the crystal. $n_S$ and $n_S$ are the corresponding refractive indices of the crystal. $A$ is the transverse cross-section. The effective free spectral range, $\Delta\omega$, of the cavity is defined as $\frac{2\pi}{T}$, with the effective cavity round-trip time $T = \frac{2l}{v_{gS}} + \frac{2(L-l)}{c}$, where $v_{gS}$ and $L$ are the group velocity of the signal field inside the PDC crystal and the resonator length, respectively. We assume that $\Delta\omega_c \cong \Delta\omega$. In Eq. (2), $\gamma$ is the cavity damping (loss) rate, $\hat{a}(\omega)$ and $\hat{b}(\omega)$ are the lowering operators of the signal and idler photons, respectively. They satisfy the following commutation relations:

$$[\hat{a}(\omega_m+\Omega),\hat{a}^{\dagger}(\omega_{m\prime}+\Omega')] = \delta_{m,m\prime}\delta(\Omega-\Omega')$$
$$[\hat{b}(\omega_1),\hat{b}^{\dagger}(\omega_2)] = \delta(\omega_1-\omega_2)$$
$$[\hat{a}(\omega_1),b^{\dagger}(\omega_2)] = 0. \tag{3}$$

In Eq. (2), the lower limits of the summation over $m$ and the integration over $\Omega$ are approximately extended to $-\infty$ because the spectral bandwidths of the signal and idler fields are substantially smaller than their center frequencies.

In the expression for $E_{S,c}^{(+)}(z,t)$ in Eq. (2), the complex spectral function $\sqrt{\gamma}/(\frac{\gamma}{2}+i\Omega)$, where its real part is related to the Lorentzian function, describes the broadening of cavity-resonant modes due to the finite damping rate $\gamma$ that is related to the cavity finesse ($F$) as $F = \pi/\gamma$. The good-cavity (e.g., weakly lossy cavity) limit corresponds to the case $\gamma \ll \Delta\omega$, meaning that the spectral bandwidth of each cavity-resonant signal mode is substantially narrower than the frequency spacing between two neighboring comb teeth of the signal field.

Inserting Eqs. (2) into (1) results in the PDC Hamiltonian:

$$\widehat{H}_{int}(t) = i\hbar\alpha \sum_{m=-\infty}^{\infty} \int_{-\infty}^{\infty} d\Omega \int_{-\infty}^{\infty} d\Omega' \frac{\sqrt{\gamma}}{(\gamma/2) - i\Omega} F_m(\Omega, \Omega')$$
$$\times \hat{a}^{\dagger}(\omega_m + \Omega')\hat{b}^{\dagger}(\omega_I + \Omega')e^{i(m\Delta\omega + \Omega + \Omega')t} + H.a., \quad (4)$$

where the constant $\alpha$ is linearly proportional to $\chi$ and determines the PDC efficiency, the biphoton production rate. $F_m(\Omega, \Omega')$ is the joint spectral function [33], and *H.a.* denotes the Hermitian adjoint. The formal solution of the time-dependent Schrödinger equation is Taylor-expanded up to the first-order term, which results in the biphoton wavefunction describing the signal-idler field at a time $t$, $|\psi(t)\rangle$. We assumed that the second-order susceptibility is weakly dependent on the signal and idler frequencies, and the phase mismatch is induced by the difference in the group velocities of the signal and idler fields in the PDC crystal. Then, considering the first-order perturbation term and approximating the *sinc* function appearing after the integration of $-i\hbar^{-1}H_{int}(t)|0\rangle$ over time with a Dirac delta function, one finds that the approximate biphoton wavefunction, which describes the state of the radiation field when exactly one signal-idler photon pair is present, can be written approximately as:

$$|\psi\rangle = \mathcal{N} \sum_{m=-\infty}^{\infty} \Phi_m \int_{-\infty}^{\infty} d\Omega \frac{\sqrt{\gamma}}{(\gamma/2) - i\Omega} \hat{a}^{\dagger}(\omega_S + m\Delta\omega + \Omega)\hat{b}^{\dagger}(\omega_I - m\Delta\omega - \Omega)|0,0\rangle_{s,i}, \quad (5)$$

where $\mathcal{N}$ is the normalization factor and the phase-mismatch factor $\Phi_m$ for the *m*th cavity-resonant signal mode is defined as

$$\Phi_m = sinc\left(\frac{m\Delta\omega\tau}{2}\right) e^{-im\Delta\omega\tau/2}. \quad (6)$$

Here, $\tau$ is the difference between the transit times of signal and idler photons through the crystal, which are determined by the corresponding group velocities, i.e., $\tau = l(v_{gI}^{-1} - v_{gS}^{-1})$. The signal and idler are initially in the vacuum state, $|0,0\rangle_{s,i}$.

The biphoton state in Eq. (5) satisfies energy conservation, so the sum of signal and idler frequencies is $\omega_P$ ($= \omega_S + \omega_I$). The spectra of both signal and idler photons (Figure 1(a)) exhibit comb structures, where each comb line has a spectral width of $\gamma$. As the damping rate gets smaller, the signal and idler single photons could have spectral bandwidths narrow enough to be used for atomic transitions.

**B. Cavity effects on the ZWM interferometer**

The ZWM interferometer is constructed by using two SPDC crystals so that the idler field from the first crystal propagates through the second crystal and is made to become indistinguishable from the idler field generated by the second SPDC process (Figure 1(b)).

There are two identical cavity-PDC systems, where each PDC can generate a signal-idler biphoton state described by Eq. (4). The first PBS on the left-hand side of the first cavity does not play a specific role except that it is needed to make the two combined cavity-PDC systems identical to each other. Furthermore, the spectra of the combined signal-idler fields of PDC-2 should be made the same as those of signal-idler fields of PDC-1 by adjusting the cavity length of one of the two.

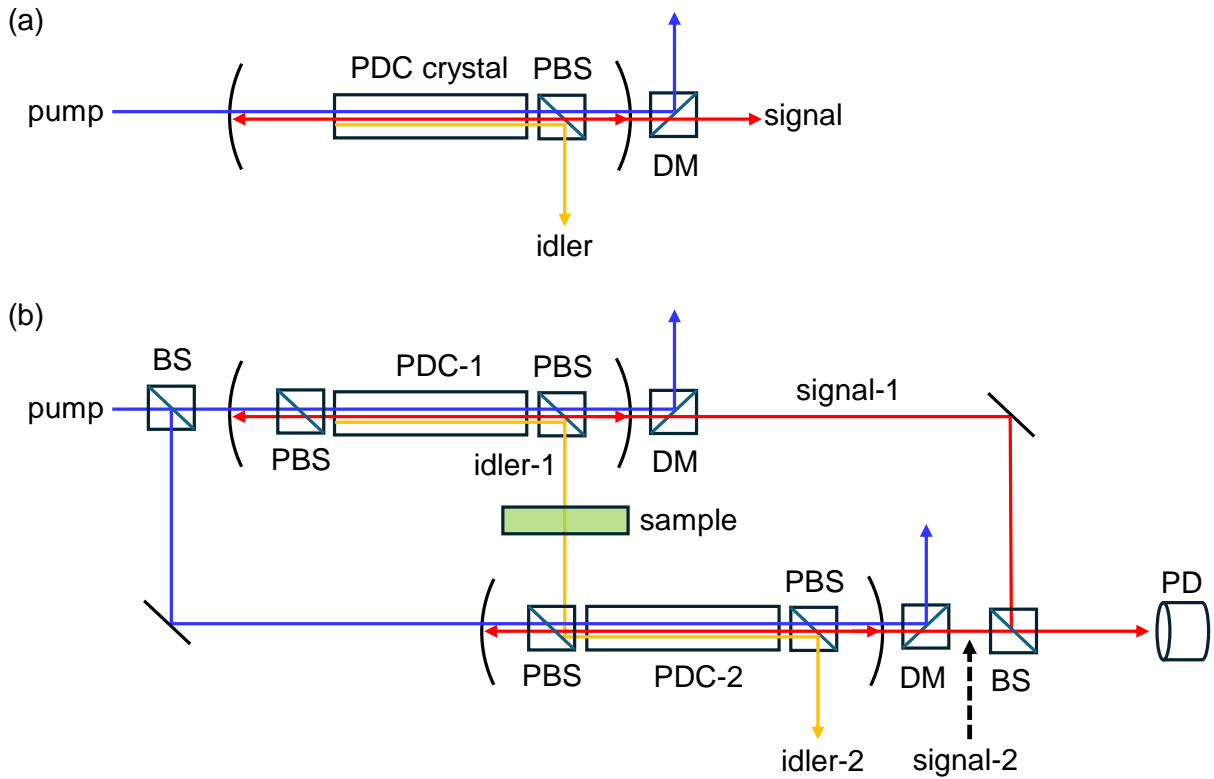

**Figure 1. Cavity-enhanced PDC and ZWM interferometer.** (a) Scheme of the cavity-enhanced PDC. The cavity is resonant with the generated signal field. PBS separates the idler field, whose polarization is orthogonal to the pump and signal, because the crystal is a type-II nonlinear optical material. (b) Scheme of the cavity-enhanced ZWM interferometer. The idler-1 field from the first PDC passes through a sample and is then injected into the second PDC crystal in another cavity. The two idler fields are aligned to be collinear with each other and become indistinguishable. The two signal fields at a single-photon level are allowed to self-interfere. (P)BS: (polarization) beam splitter, PDC: parametric down-conversion, DM: dichroic mirror, PD: photo-detector.

Due to the phase-matching conditions, there is only one cavity-resonant signal mode that is frequency-correlated with $\omega_{i_1} = \omega_P - \omega_{s_1}$. An interesting point is the presence of a sample, considered a lossless BS, between the two cavity-PDC systems. As can be seen in Figure 1(b), the idler-1 field interacts with a sample that could absorb idler-1 photons from the PDC-1 in Figure 1(b). The amplitude transmissivity, denoted as $T(\omega_{i_1})$, of the sample, which is assumed to be any four-port device with two input and two output ports, is idler frequency-dependent. Here, the idler-1 is considered to be normally incident on the front and rear surfaces of the sample. Then, when the idler-1 and idler-2 beams are perfectly aligned in a collinear geometry, the annihilation operator $\hat{b}_2$ of idler-2 (or the amplitude of the idler-2 field) can be related to $\hat{b}_1$ of idler-1 as

$$\hat{b}_2(\omega_i) = [T(\omega_i)\hat{b}_1(\omega_i) + R'(\omega_i)\hat{b}_0(\omega_i)]e^{i\varphi}, \tag{7}$$

where $R'$ is the amplitude reflectivity of the rear-end (bottom) surface of the sample at the idler frequency. $\hat{b}_0$ is the annihilation operator of the vacuum field at the unused port of the sample, which propagates in the direction opposite to the idler-1 field. In Eq. (7), $\varphi$ is the phase accrued by the idler field during its propagation from PDC-1 to PDC-2, which is assumed to be frequency-independent within the spectral bandwidth, $2\pi/|\tau|$, due to the phase mismatch.

Now, using the Hamiltonian in Eq. (4) and invoking the same line of approximations used to obtain Eq. (5), we obtain the biphoton wavefunction of the cavity-enhanced ZWM interferometer:

$$|\psi(t)\rangle = \mathcal{N} \sum_{m=-\infty}^{\infty} \Phi_m \int_{-\infty}^{\infty} d\Omega \frac{\sqrt{\gamma}}{(\gamma/2) - i\Omega} \{\hat{a}_1^\dagger(\omega_S + m\Delta\omega + \Omega)\hat{b}_1^\dagger(\omega_I - m\Delta\omega - \Omega)$$

$$+ \hat{a}_2^\dagger(\omega_S + m\Delta\omega + \Omega)[T^*(\omega_I - m\Delta\omega - \Omega)\hat{b}_1^\dagger(\omega_I - m\Delta\omega - \Omega)$$

$$+ R'^*(\omega_I - m\Delta\omega - \Omega)\hat{b}_0^\dagger(\omega_I - m\Delta\omega - \Omega)]e^{i\varphi}\}|0,0\rangle_{s_1,s_2}|0,0\rangle_{i_1,i_0}, \tag{8}$$

where $\hat{a}_1^\dagger$ and $\hat{a}_2^\dagger$ are the creation operators of the signal-1 and signal-2 photons, respectively. $|0,0\rangle_{i_1,i_0}$ represents the vacuum state of the idler-1 and idler-0 fields. The resulting biphoton state is a superposition of biphoton states prepared by the upper and lower crystals. The tensor product of two biphoton states can be safely ignored within the limit of low PDC efficiency.

If the amplitude transmissivity $T$ of the sample is unity, the indistinguishability of the two idler fields becomes maximal, which makes the path (origin) of a generated signal single-photon indeterminable. On the other hand, in the limit $T = 0$,
if a detector is placed at the output port of idler-2 and detects single idler photons, these photons should originate from PDC-2, which suggests that the idler photons are distinguishable. Consequently, the path of a signal photon becomes determinable in this case. To examine the induced coherence of signal photons, which is modulated by the indistinguishability of frequency-entangled idler photons, one can measure the frequency-resolved visibility of the signal field at the photo-detector.

The output spectrum of the signal field is given by the Fourier transform of the first-order temporal correlation function, i.e., the expectation value of $\hat{E}_S^{(-)}(z,t)\hat{E}_S^{(+)}(z,t+t')$ calculated with the biphoton wavefunction in Eq. (8). Here, $\hat{E}_S^{(+)}(z,t)(\hat{E}_S^{(-)}(z,t))$ is the positive-(negative-)frequency component of the signal electric field in free space at the PD in Figure 1(b), which is given by:

$$\hat{E}_S^{(+)}(z,t) = \sqrt{\frac{\hbar \omega_S}{2\epsilon_0 cA}} \int_{-\infty}^{\infty} \frac{d\Omega}{\sqrt{2\pi}} [\hat{a}_1(\omega_S + \Omega)e^{i(\omega_S+\Omega)(\frac{z}{c}-t)} + \hat{a}_2(\omega_S + \Omega)e^{i(\omega_S+\Omega)(\frac{z}{c}-t)+i\phi}], \quad (9)$$

where $\phi$ is the relative phase factor between the signal-1 and signal-2 fields experiencing different dispersive properties of the BS and other optical components. $\phi$ is assumed to be a frequency-independent constant because the spectral bandwidth, $\sim 2\pi/|\tau|$, due to the phase mismatch, is narrow.

Using the commutation relationships in Eq. (3) and carrying out the Fourier transformation of the time-correlation function, we obtain

$$S(\omega) \propto \sum_{m',m=-\infty}^{\infty} \mathrm{sinc}\left(\frac{m'\Delta\omega\tau}{2}\right) \mathrm{sinc}\left(\frac{m\Delta\omega\tau}{2}\right) e^{i(m'-m)\Delta\omega\tau/2}$$

$$\times \frac{\gamma}{[(\frac{\gamma}{2}) + i(\omega - \omega_S - m'\Delta\omega)][(\frac{\gamma}{2}) - i(\omega - \omega_S - m\Delta\omega)]}$$

$$\times [1 + |T(\omega_P - \omega)|^2 + 2|T(\omega_P - \omega)|\cos(\phi + \varphi)]. \quad (10)$$

In this paper, we consider the limiting case of two cavity-enhanced optical parametric oscillators below the threshold and assume that the cavity damping rate is substantially lower than the free spectral range, i.e., $\gamma \ll \Delta\omega$. Furthermore, the difference between the transit times of the signal and idler photons through a crystal is small compared to the cavity round-trip time of a signal photon, i.e., $\Delta\omega \ll 1/|\tau|$. Therefore, Eq. (10) can be approximated as

$$S(\omega) \propto \sum_{m=-\infty}^{\infty} \mathrm{sinc}^2\left(\frac{m\Delta\omega\tau}{2}\right) \frac{\gamma}{(\gamma/2)^2 + (\omega - \omega_S - m\Delta\omega)^2}$$

$$\times [1 + |T(\omega_P - \omega)|^2 + 2|T(\omega_P - \omega)|\cos(\phi + \varphi)]. \quad (11)$$

One can further assume that the sample's transmissivity $T(\omega)$ is, within the spectral bandwidth determined by $\gamma$, a weakly varying function of frequency at around $\omega = \omega_S - m\Delta\omega$. Then, we can simplify Eq. (11) as

$$S(\omega) \propto \sum_{m=-\infty}^{\infty} \mathrm{sinc}^2\left(\frac{m\Delta\omega\tau}{2}\right) \frac{\gamma}{(\gamma/2)^2 + (\omega - \omega_S - m\Delta\omega)^2}$$

$$\times [1 + |T(\omega_I - m\Delta\omega)|^2 + 2|T(\omega_I - m\Delta\omega)|\cos(\phi + \varphi)]. \quad (12)$$

From Eq. (12), it is clear that the spectrum of signal photons is determined by four factors: (i) the sinc function resulting from phase mismatch, (ii) the cavity-resonant signal mode frequencies ($\omega_S + m\Delta\omega$), (iii) the cavity damping rate, $\gamma$, of the signal field, and (iv) the amplitude transmissivity of the sample at the phase-matched idler frequency ($\omega_I - m\Delta\omega$). Furthermore, the intensity at the peak of one of the output signal modes is enhanced by the effect of cavity resonance of the signal field, which is determined by its finesse.

Assuming there is a comb-resolving spectrometer, one can measure the fringe visibility associated with signal photons whose frequencies are within the Lorentzian band at $\omega = \omega_S - m\Delta\omega$, to examine the extent of self-coherence as in a single-photon double-slit experiment. The visibility of the $m$th signal comb mode, from Eq. (12), is given by

$$V_m = \frac{2|T(\omega_I - m\Delta\omega)|}{1 + |T(\omega_I - m\Delta\omega)|^2}. \qquad (13)$$

Suppose the sample consists of a mixture of atoms with different absorption frequencies but whose frequencies are in the envelope of the spectral bandwidth ($\sim 2\pi/|\tau|$) determined by the phase mismatch. In that case, one can measure the atomic transition-resolved spectrum of such an atomic ensemble using this cavity-enhanced ZWM interferometer employing narrow-band idler photons that are frequency-entangled with the signal photons.

## 3. SUMMARY

We introduced a theoretical framework for cavity-enhanced induced coherence without induced emission, where each OPO is assumed to operate below the threshold. The cavity effect on SPDC was shown to be useful for producing signal-idler single photons with a linewidth on the order of cavity damping rate, which is required to interface with the narrow atomic transitions in quantum communications and memories. We showed that the fringe visibility of signal single-photon fields is modulated by the indistinguishability of frequency-entangled idler comb modes whose linewidth is determined by the cavity damping rate. We suggest that this cavity-enhanced ZWM interferometry could be used in comb-resolved quantum spectroscopy and metrology with undetected photons.

**Funding**

Institute for Basic Science for financial support (IBS-R023-D1 for MC).

## Declaration of competing interest

The authors declare that they have no known competing financial interests or personal relationships that could have appeared to influence the work reported in this paper.

## Data availability

Data will be made available on request.